\documentclass[useAMS,usenatbib]{mn2e}

\def\degr{\hbox{$^\circ$}}
\def\arcmin{\hbox{$^\prime$}}
\def\arcsec{\hbox{$^{\prime\prime}$}}
\def\flx{erg cm$^{-2}$ s$^{-1}$}
\def\lum{erg s$^{-1}$}

\def\chan{{\it Chandra}}

\usepackage{epsfig}
\usepackage{graphicx}
\usepackage{times}
\usepackage{float}
\usepackage{color}
\usepackage{aas_macros}

\title[Two fast X-ray transients in archival Chandra data]{Two fast X-ray transients in archival Chandra data}
\author[A. Glennie, P. G. Jonker, R. P. Fender, T. Nagayama and M. L. Pretorius]{A. Glennie$^{1}$\thanks{E-mail: aidan.glennie@astro.ox.ac.uk}, P. G. Jonker$^{2,3,4}$, R. P. Fender$^{1}$, T. Nagayama$^{5,6}$ and M. L. Pretorius$^{1}$\\
$^{1}$Department of Physics, University of Oxford, Oxford, OX1 3RH, UK\\
$^{2}$SRON, Netherlands Institute for Space Research, Sorbon-nelaan 2, 3584 CA, Utrecht, The Netherlands\\
$^{3}$Department of Astrophysics/IMAPP, Radboud University Nijmegen, P.O.~Box 9010, 6500 GL, Nijmegen, The Netherlands\\
$^{4}$Harvard--Smithsonian  Center for Astrophysics, 60 Garden Street, Cambridge, MA~02138, USA\\
$^{5}$Department of Astrophysics, Nagoya University, Furo-cho, Chikusa-ku, Nagoya, Aichi 464-8602, Japan\\
$^{6}$Department of Science and Engineering, Kagoshima University, Korimoto, Kagoshima 890-0065, Japan}
\begin{document}

\date{\today}

\pagerange{\pageref{firstpage}--\pageref{lastpage}} \pubyear{2014}

\maketitle

\begin{abstract}
  We present the discovery of two new X-ray transients in archival \chan\, data. The first transient, XRT110103, occurred in January 2011 and shows a sharp rise of at least three orders of magnitude in count rate in less than 10 s, a flat peak for about 20 s and decays by two orders of magnitude in the next 60 s. We find no optical or infrared counterpart to this event in preexisting survey data or in an observation taken by the SIRIUS instrument at the Infrared Survey Facility $\sim2.1$ yr after the transient, providing limiting magnitudes of $J>18.1$, $H>17.6$ and $K_s>16.3$. This event shows similarities to the transient previously reported in Jonker et al. which was interpreted as the possible tidal disruption of a white dwarf by an intermediate mass black hole. We discuss the possibility that these transients originate from the same type of event. If we assume these events are related a rough estimate of the rates gives $1.4\times10^{5}$ per year over the whole sky with a peak 0.3 - 7 keV X-ray flux greater than $2\times10^{-10}$ \flx. The second transient, XRT120830, occurred in August 2012 and shows a rise of at least three orders of magnitude in count rate and a subsequent decay of around one order of magnitude all within 10 s, followed by a slower quasi-exponential decay over the remaining 30 ks of the observation. We detect a likely infrared counterpart with magnitudes $J=16.70\pm0.06$,  $H=15.92\pm0.04$ and $K_s=15.37\pm0.06$ which shows an average proper motion of $74\pm19$ milliarcsec per year compared to archival 2MASS observations. The $JHK_s$ magnitudes, proper motion and X-ray flux  of XRT~120830 are consistent with a bright flare from a nearby late M or early L dwarf.
\end{abstract}


\section{Introduction}

Astronomical transients represent huge releases of energy, often on very short timescales, and their discovery and follow-up allows us to probe the extreme physics involved. X-rays, being associated in large part with compact, high energy density processes such as accretion, can reveal much of the astrophysics in such events. The brightest X-ray sources in the sky are associated with accreting neutron stars and black holes and the vast majority are highly variable.

Transients which can be detected as variable within individual \chan\ observations would be variable on timescales of seconds up to tens of kiloseconds for the longest observations. Within our galaxy the major populations of X-ray transients exhibiting bursting behaviour on time-scales of seconds to hours come from flare stars and X-ray binary systems. Flare stars can produce X-ray bursts on time-scales of minutes to hours due to magnetic reconnection events in the stellar atmosphere \citep{osten05}. Accretion events in the bulk of X-ray binaries can show variability on these timescales, but in general with no more than $\sim50$\% amplitude \citep[e.g.][]{lewin06}. Type 1 X-ray bursts are observed in neutron star X-ray binaries and are caused by thermonuclear flashes from accreted material on the neutron star surface \citep[e.g.][]{strohmayer06,galloway08}. These occur on time-scales of a few to hundreds of seconds releasing total energy of around $10^{39} - 10^{40}$ ergs. Supergiant fast X-ray transients (SFXTs) are high mass X-ray binary systems with supergiant companions. SFXTs can produce X-ray flares on typical time-scales of 100-10,000 s \citep{sidoli13}.

Extragalactic populations of X-ray transients such as X-ray rich gamma-ray bursts and X-ray flashes \citep{sakamoto08} can produce rapid bursts of X-rays with a wide range of peak energies. Typically tidal disruption events evolve over too long a timescale \citep[months to years, e.g.][]{komossa04} to be detected as variable over a single \chan\ observation but the time-scale of the event decreases with black hole mass, radius of the disrupted star and the impact factor (beta) \citep[e.g. equation 4 of][]{lodato11}. This suggests the possibility of faster X-ray transients from the tidal disruption of compact stars around intermediate mass black holes \citep[e.g. in the case of white dwarfs see][]{rosswog09}.

There have recently been two discoveries of \chan\ transients on time-scales of hundreds of seconds with probable extragalactic origin. \citet{jonker13} report a transient close to M86 with possible precursor events and a peak luminosity $\sim6\times10^{42}$ \lum (assuming the distance of M86). This was interpreted as the possible tidal disruption of a white dwarf by an intermediate mass black hole, although other explanations such as an off-axis GRB are not ruled out. \citet{atel6541} report a transient in the Chandra Deep Field-South survey of duration $\sim1$ ks possibly associated with a R=27.38 (AB) galaxy with photometric redshift $0.31\pm0.16$ \citep{atel6603}. 

In this paper we report the discovery of two faint X-ray transients in archival \chan\ data. The analysis is split into two parts covering each transient discovered in turn. Follow up near-infrared and archival Very Large Array (VLA), Two Micron All Sky Survey (2MASS) and United States Naval Observatory (USNO-B) observations of the field are analysed in a search for possible counterparts. We also analyse the spectral evolution of the transient discovered by \citet{jonker13}, XRT~000519. We then discuss the probable nature of these transients including the possibility that one is of similar nature to those discovered recently by \citet{jonker13} and \citet{atel6541}.

\section{Observations and analysis}

We have searched the Chandra Data Archive for rapid transient events, making use of the Chandra Source Catalog \citep{evans10} and conducting our own search of data publicly released from 2010 onwards (Glennie et al. in preparation).

There is a rich sample of statistically variable sources in the Chandra Data Archive. We present two of the more unique transient sources discovered, which appear to be relatively bright and evolve over a short timescale. The first was discovered in an observation of the galaxy cluster ACO 3581 with observation identification number (ObsID) 12884 which began on 2011 January 3. We name this transient after the date of the event, XRT~110103. The observation in which the second transient was discovered was targeting the galaxy cluster SPT-CLJ2352-4657 with ObsID 13506 beginning 2012 August 30. We name this transient XRT~120830. 

The details of the \chan\ observations and source coordinates are listed in Table \ref{obs}. All errors given are to the $1\sigma$ level of significance unless otherwise stated. The \chan\ data processing was done using {\sc CIAO 4.5} and {\sc CALDB 4.5.6}. The source positions were derived using the {\sc CIAO} tool {\sc wavdetect}. Source photons for each transient were extracted from a circular region centred on the {\sc wavdetect} position with radii of 30\arcsec. This relatively large extraction region is warranted by the large off-axis angle of the transients of 13.3 and 12.3 arcmin for XRT~110103 and XRT~120830 respectively. In each case using an annulus as the background region would extend beyond the edge of the ACIS CCD so a nearby region of radius 60\arcsec was chosen, centred on RA \& DEC $14^{\rmn{h}} 08^{\rmn{m}} 30^{\rmn{s}} -27\degr 05\arcmin 28\arcsec$ for XRT~110103 and $23^{\rmn{h}} 52^{\rmn{m}} 11^{\rmn{s}} -46\degr 41\arcmin 47\arcsec$ for XRT~120830. The tool {\sc specextract} was used to extract the source spectra and {\sc Xspec 12.8.0} used for spectral fitting. In each case source photons in the range 0.3 - 7.0 keV were used for spectral analysis and the Cash statistics \citep{cash79} used to determine the best-fit model parameters.

Follow up near-infrared observations with the SIRIUS instrument at the Infrared Survey Facility \citep[IRSF,][]{nagayama03} were analysed using {\sc Starlink GAIA 4.4.4}. Archival VLA data of the field were reduced using the VLA Calibration Pipeline 1.2.0 and imaging and analysis performed with {\sc CASA 4.1.0}.

\begin{table}
\centering
\caption{Details of the observations and coordinates for each transient. The source position uncertainties are estimated using the method set out in \citet{evans10} including the ChaMP positional uncertainty relations \citep{kim07} and the astrometric error calculated by \citet{rots11}.}
\begin{tabular}{cccc}
\hline
Transient & ObsID & RA \& DEC J2000 & Error \\
\hline
XRT~110103 & 12884 & $14^{\rmn{h}} 08^{\rmn{m}} 28\fs89 -27\degr 03\arcmin 29\farcs4$ & 1\farcs1 \\
XRT~120830 & 13506 & $23^{\rmn{h}} 52^{\rmn{m}} 12\fs19 -46\degr 43\arcmin 43\farcs3$ & 1\farcs5 \\
\hline
\end{tabular}
\label{obs}
\end{table}

\subsection{XRT~110103}

\begin{figure}
\centering
\includegraphics[width=0.5\textwidth]{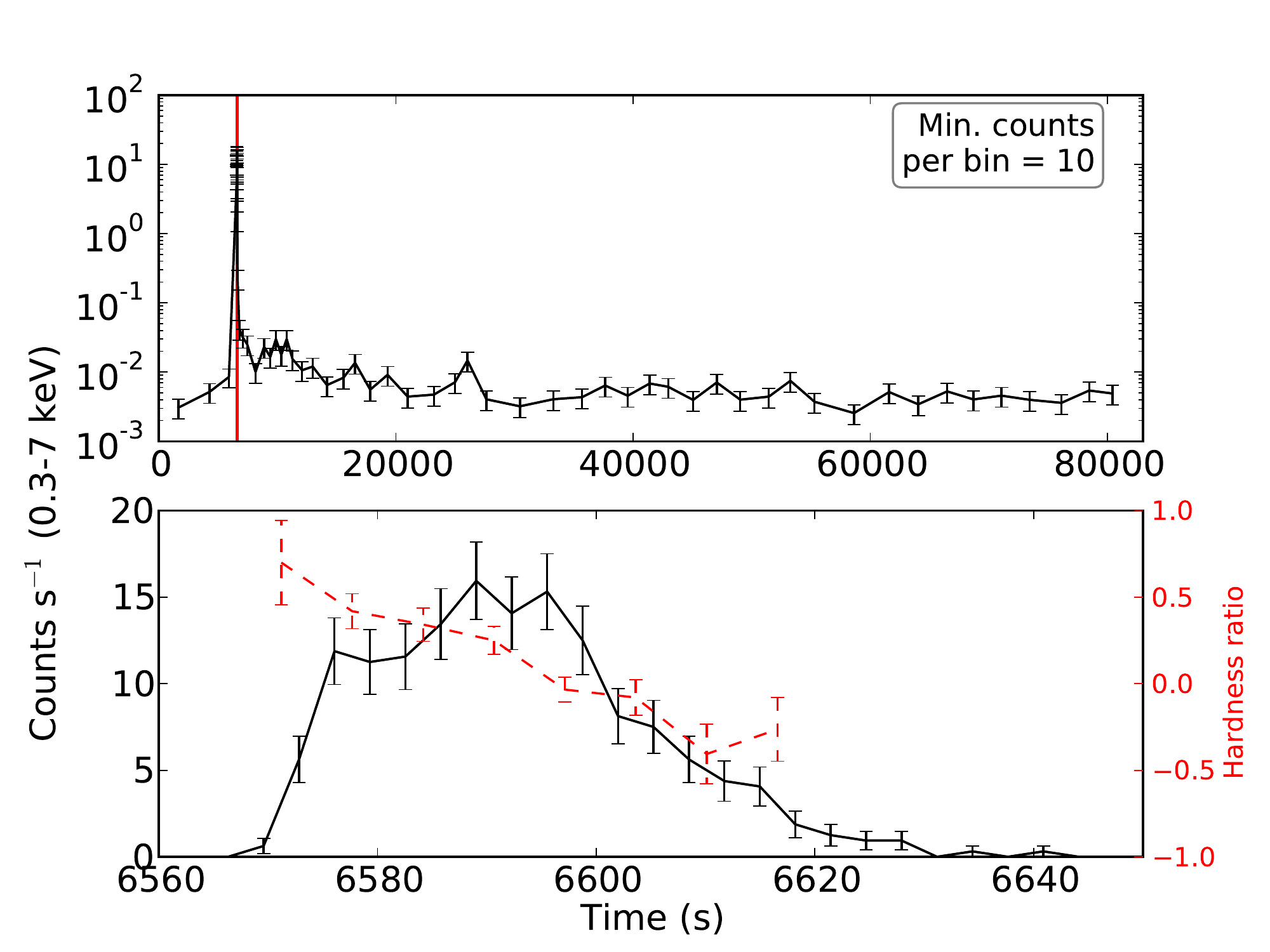}
\caption{Lightcurves of XRT~110103 showing the variability over the entire \chan\, observation (top panel, logarithmic count rate) and a close-up of the main flare (bottom panel, linear count rate). The vertical line on the top panel indicates the time period shown in the bottom panel. The dashed (red) line indicates the evolution of spectral hardness (axis on the right of the bottom panel). The time resolution of the bottom panel lightcurve is 3.24 s and the hardness ratio is 6.48 s. The source is not detected over a background count rate of $4.1 \times 10^{-3}$ counts s\textsuperscript{-1} prior to the main flare (the lightcurves are not background subtracted).}
\label{lc1}
\end{figure}

The evolution of the count rate as a function of time of XRT~110103 is shown in Figure \ref{lc1}. The lightcurves were extracted with the maximum time resolution for ACIS of 3.24 s. The top panel shows the evolution of the transient over the entire observation. In the periods of the observation where we obtain a low count rate we average time bins together to ensure a minimum of 10 counts per bin. The vertical line on the top panel indicates the time period shown in the bottom panel to illustrate the features of the main flare. In the 6560s of the observation prior to the main flare we do not significantly detect the source above the background count rate of $4.1 \times 10^{-3}$ counts s\textsuperscript{-1}. During the main flare the count rate increases by more than 3 orders of magnitude in less than 10 s, there is a flat peak for around 20 s and then a steady decay of around 2 orders of magnitude in the next 60 s. We detect some spectral softening over the duration of the flare, illustrated with the hardness ratio in Figure \ref{lc1}. We have defined the hardness ratio as follows and used 6.48 s time bins to maximise signal to noise.

\[
 Hardness = \frac{F_h - F_s}{F_s + F_m + F_h}
\]

Where $F_s$, $F_m$ and $F_h$ are the photon fluxes in the soft (0.5 - 1.2 keV), medium (1.2 - 2 keV) and hard (2 - 7 keV) bands. After the main flare the source follows a more gradual decay for $\sim$20 ks after which it is not significantly detected above the background count rate for the remainder of the observation.

\begin{figure}
\centering
\includegraphics[width=0.4\textwidth]{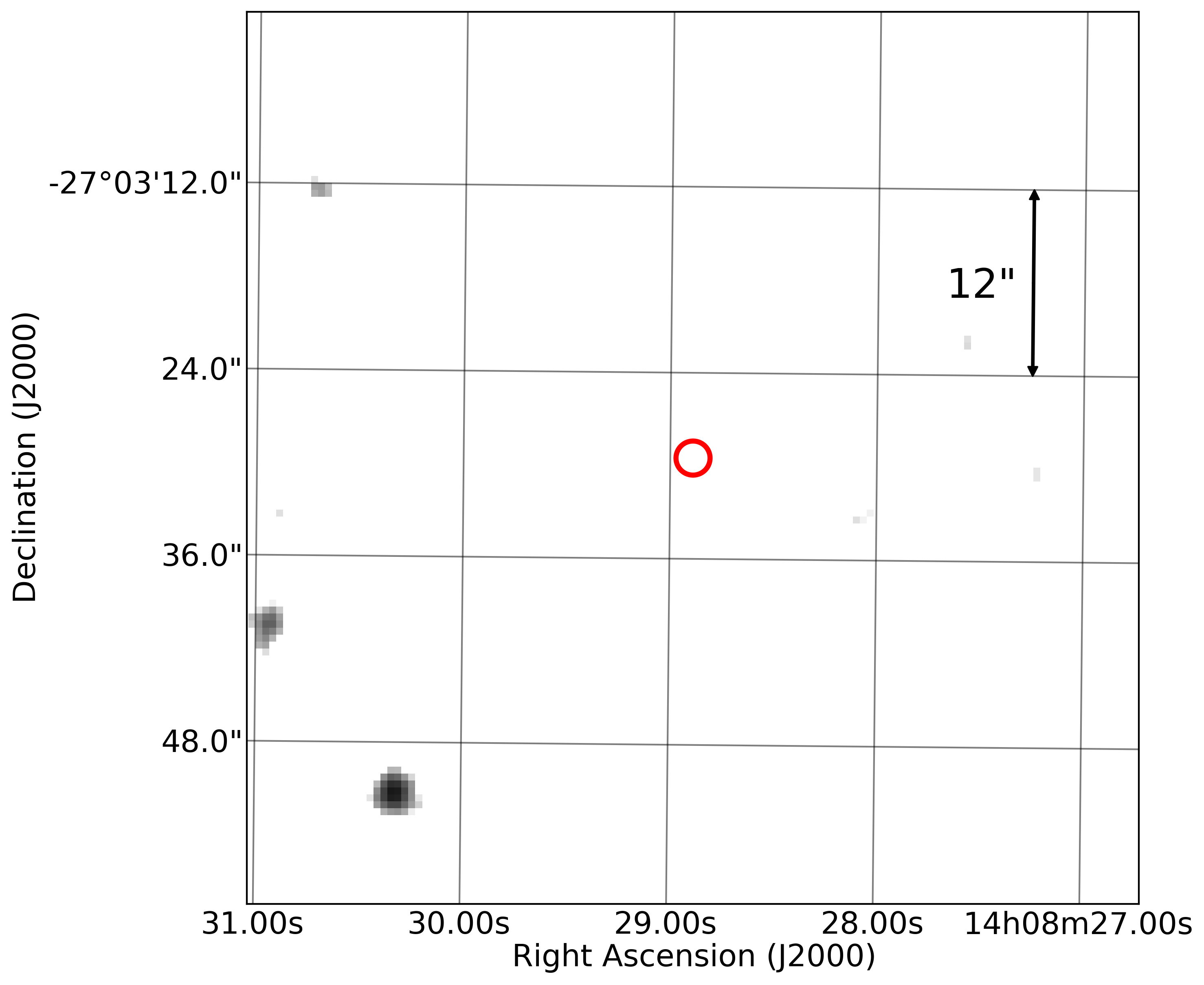}
\caption{This IRSF/SIRIUS {\it J}-band image shows the $1\sigma$ \chan\ position of XRT~110103 is not consistent with any infrared source we detect. We have used a Gaussian smoothing with a kernel radius of 1 pixel.}
\label{abel3581}
\end{figure}

XRT~110103 was discovered in an observation of the galaxy cluster ACO 3581. However, the nearest known member of the galaxy cluster, ACO 3581 12, is 2.7 arcmin from the \chan\ transient position. We obtained a follow up observation with the IRSF on 2014 March 14, $\sim2.1$ yr after the transient detection. Details of the observation are shown in Table \ref{irsf_obs}. Figure \ref{abel3581} shows the J-band image from this observation. There is no counterpart within the \chan\ error circle, placing limiting magnitudes ($5\sigma$) of {\it J}$>$18.1, {\it H}$>$17.6 and {\it K$_s$}$>$16.3. We find no optical counterpart in the USBO-B catalog \citep{monet03} implying a limit of R$>$20.9. We also analysed an archival L-band (1.34 - 1.73 GHz) VLA observation of the field taken $\sim1.4$ yr after the transient (project code SB0410 starting on 2012 May 11). We found no counterpart down to an upper limit of 0.25 mJy beam\textsuperscript{-1} (three times the root-mean-square noise at the position of the transient).

We have fitted an absorbed powerlaw model (pegpwrlw in Xspec) to the X-ray spectrum of the main burst with various values for the absorption. A total of 470 source photons were taken from the 60 s following the start of the transient event as shown in the bottom panel of Figure \ref{lc1}. The results of these fits are shown in Table \ref{t2_spec}. A fit was calculated with no absorption, another with the expected line of sight Galactic foreground value of N$_H = 4.3 \times 10^{20}$ \citep{dickey90} and finally leaving the N$_H$ parameter free to find the best fit.

\begin{table}
\centering
\caption{Summary of the follow up observations taken with IRSF/SIRIUS.}
\begin{tabular}{ccccc}
\hline
Target & Telescope & Date & Seeing & Exposure (s) \\
\hline
XRT~110103 & IRSF/SIRIUS & 14 March 2014 & $2\farcs4$ & 1500 \\
XRT~120830 & IRSF/SIRIUS & 8 December 2013 & $1\farcs8$ & 1380 \\
\hline
\end{tabular}
\label{irsf_obs}
\end{table}

\begin{table}
\centering
\caption{A powerlaw model was fit to the X-ray spectrum of XRT~110103 with various corrections for absorption due to the column density, N$_H$. The value of N$_H = 4.3 \times 10^{20}$cm\textsuperscript{-2} is the expected absorption due to the galactic foreground \citep{dickey90}. The value of N$_H = 5.5 \times 10^{21}$cm\textsuperscript{-2} was found by leaving the N$_H$ parameter free to converge on the best fit.}
\begin{tabular}{ccc}
\hline
N$_H$ (cm\textsuperscript{-2}) & Peak Flux (\flx) & Photon Index \\
\hline
$0$ & $(1.1\pm0.1)\times 10^{-10}$ & $1.3\pm0.1$ \\
$4.3 \times 10^{20}$ & $(1.2\pm0.1)\times 10^{-10}$ & $1.4\pm0.1$ \\
$(5.5\pm1) \times 10^{21}$ & $(2.2\pm0.4)\times 10^{-10}$ & $2.3\pm0.2$ \\
\hline
\end{tabular}
\label{t2_spec}
\end{table}

\subsection{XRT~000519}

\begin{figure}
\centering
\includegraphics[width=0.5\textwidth]{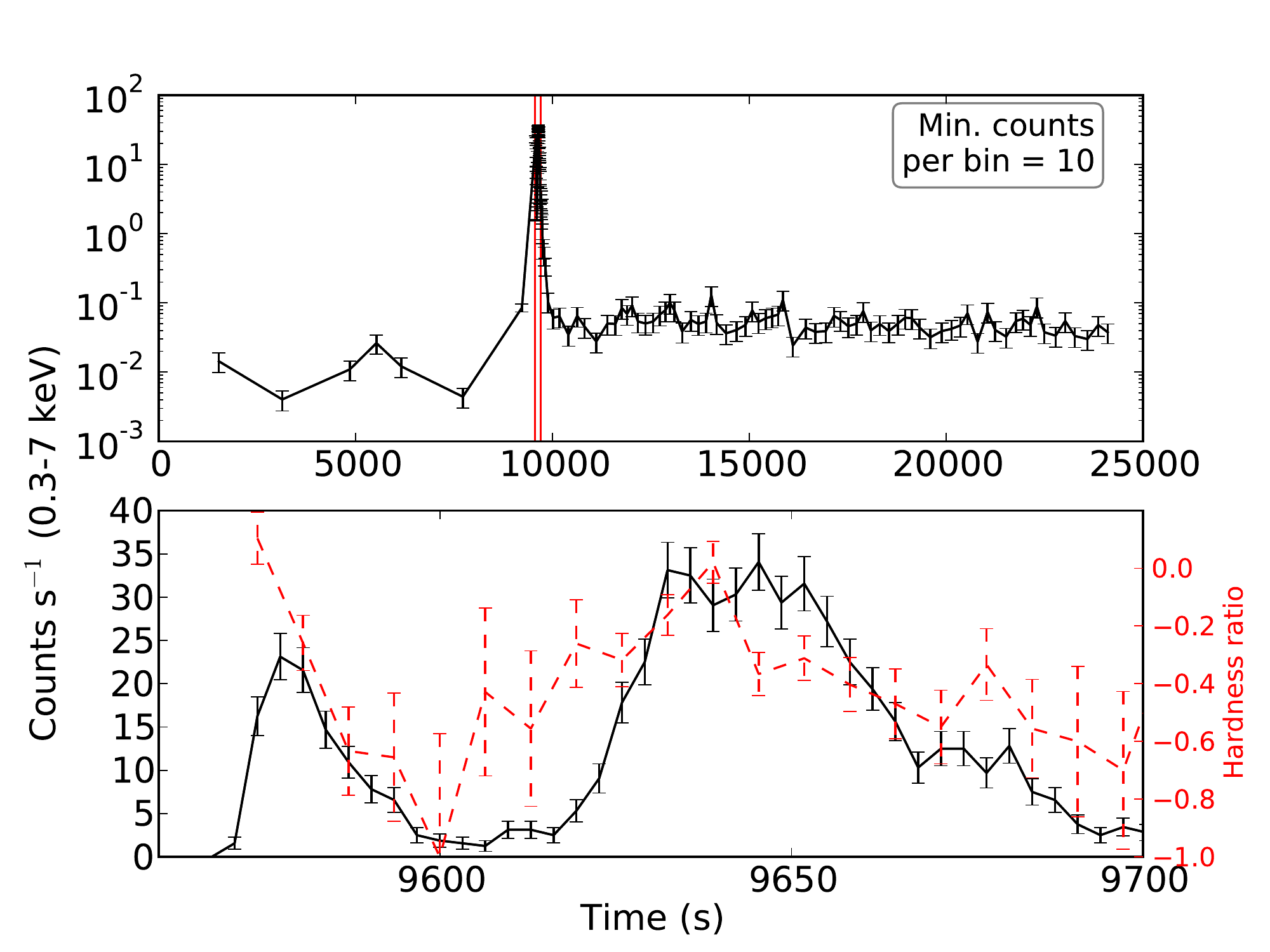}
\caption{Lightcurves of XRT~000519 showing the variability over the entire \chan\, observation (top panel, logarithmic count rate) and a close-up of the main flare (bottom panel, linear count rate). The vertical lines on the top panel indicate the time period shown in the bottom panel. The dashed (red) line indicates the evolution of spectral hardness (axis on the right of the bottom panel). The time resolution of the bottom panel lightcurve is 3.24 s and the hardness ratio is 6.48 s.}
\label{lcnew}
\end{figure}

XRT~000519 is a transient originally discovered by \citet{jonker13}, also in archival \chan\ data. It comes from a direction close to M86 and evolves over a  similar timescale to XRT~110103. XRT~000519 shows some possible precursor events $\sim4$ and $\sim8$ ks prior to the main flare. \citet{jonker13} found some spectral softening between the first and second bursts but did not display the spectral evolution of the source at greater time resolution. We have applied the same hardness ratio as for XRT~110103 and provide in Figure \ref{lcnew} an equivalent plot of the lightcurve and hardness ratio for comparison.

\subsection{XRT~120830}

\begin{figure}
\centering
\includegraphics[width=0.5\textwidth]{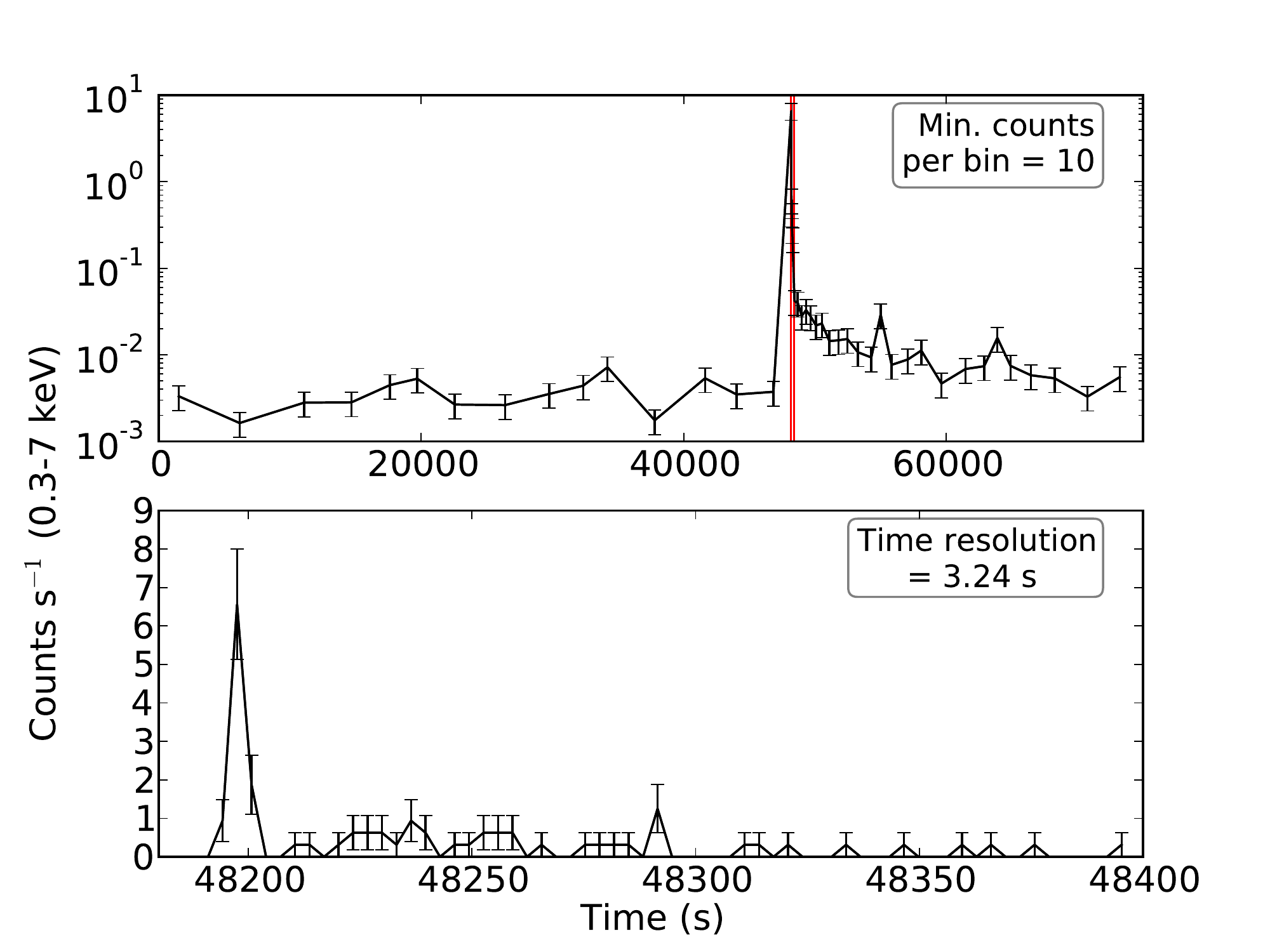}
\caption{Lightcurves of XRT~120830 showing the variability over the entire \chan\, observation (top panel, logarithmic count rate) and a close-up of the main flare (bottom panel, linear count rate) with 3.24 s time resolution. The vertical line on the top panel indicates the time period shown in the bottom panel. The source is not detected over a background count rate of $2.5 \times 10^{-3}$ counts s\textsuperscript{-1} prior to the main flare (the lightcurves are not background subtracted).}
\label{lc2}
\end{figure}

For XRT~120830 we show in Figure \ref{lc2} the lightcurve for the whole \chan\, observation. The lightcurve was extracted with 3.24 s time resolution with bins combined at low count rate periods to ensure a minimum number of 10 counts in each bin. We provide a close up of the main flare in the bottom panel to indicate the rapid rise and fall in count rate. XRT~120830 is not significantly detected prior to the main burst. We are unable to fully resolve the rise time of the main flare with the 3.24 s time resolution of ACIS. The count rate increases by 3 orders of magnitude from the background level and then rapidly decays by more than an order of magnitude from the peak count rate. This rise and decay occurs in approximately 10 s. The transient continues to decay for the remainder of the observation, with some marginally significant flaring events $\sim7$ and $14$ ks after the main burst.

\begin{figure}
\centering
\includegraphics[width=0.5\textwidth]{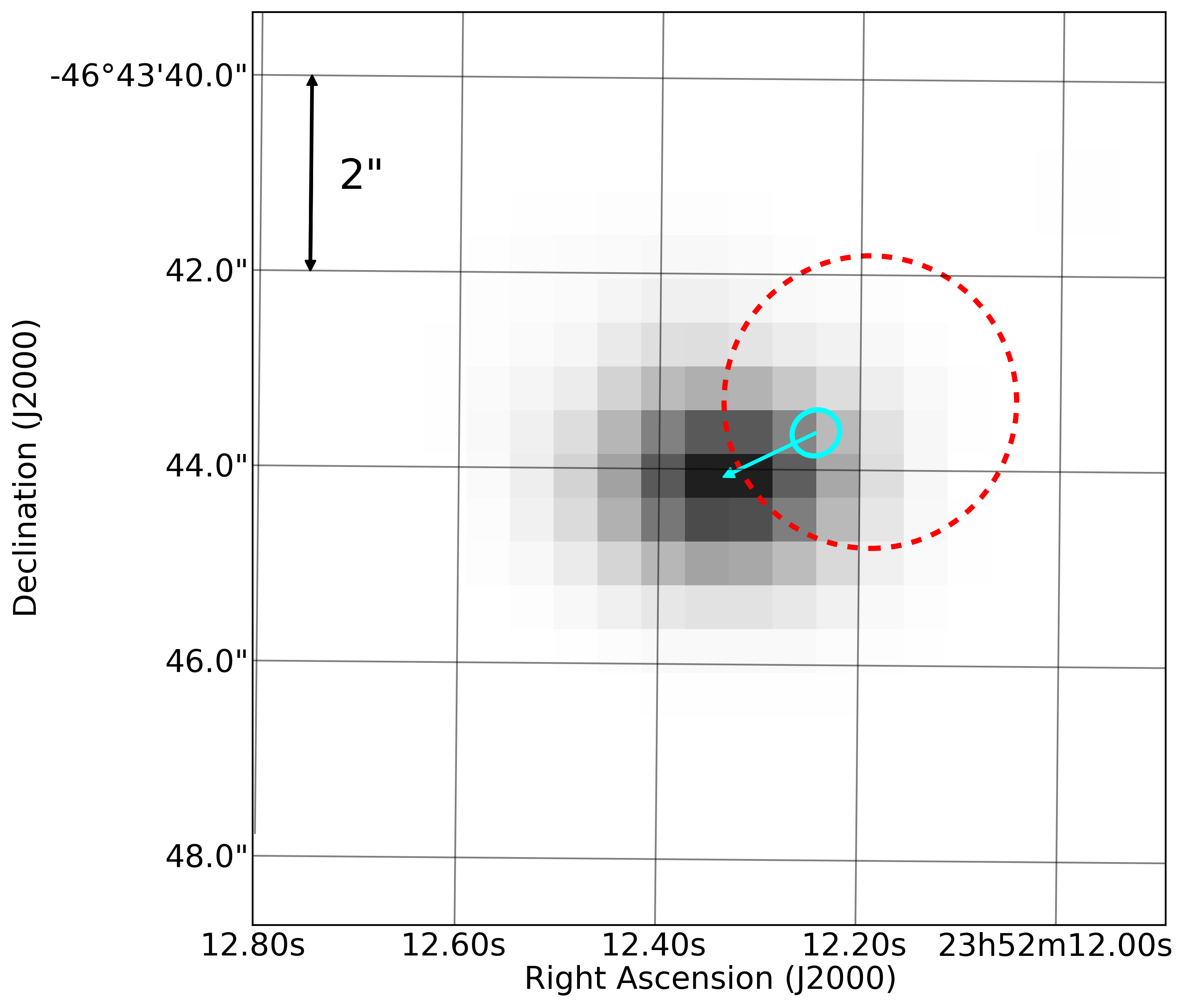}
\caption{This {\it J}-band image from IRSF/SIRIUS shows the candidate infrared counterpart to XRT~120830. We have used a Gaussian smoothing with a kernel radius of 1 pixel. The 2MASS error region is indicated with a solid blue cirlce. The projected motion between the 2MASS observation (1999 September 20) and the IRSF/SIRIUS observation (2013 December 8) is indicated with an arrow. The error circle from the \chan\ observation is displayed as a red dashed circle. The date of the \chan\ observation was 2012 August 30.}
\label{irsf}
\end{figure}

We found a possible faint infrared counterpart to XRT~120830 in the 2MASS catalog \citep{skrutskie06}. We followed this up with an observation with the Infra-Red Survey Facility (IRSF) on 2013 December 8, $\sim1.3$ yr after the transient was detected. A candidate counterpart was detected in the IRSF observation and was found not to have varied in the {\it J}, {\it H} and {\it K$_s$} bands from the 2MASS survey (see Table \ref{irsf_mag}). The position is also consistent with a Wide-field Infrared Survey Explorer \citep{cutri12} point source with magnitudes {\it W1}=15.108$\pm$0.035 {\it W2}=15.133$\pm$0.076. We find no optical counterpart in the USBO-B catalog \citep{monet03} implying a limit of {\it R}$>$20.1. We find an offset of $1\farcs05$ between the 2MASS and IRSF/SIRIUS positions. Taking the $0\farcs25$ uncertainty from the 2MASS catalog and assuming a typical $0\farcs1$ uncertainty for the IRSF detection \citep{kato07} we find a proper motion of $74\pm19$ milliarcsec per year. Given the dates of the 2MASS and IRSF/SIRIUS observations we estimate the source would need to be within $\sim10$ pc for parallax to contribute significantly to the uncertainty in this calculation. We suggest in the discussion that a distance of around 80 pc is more likely. This offset is illustrated in Figure \ref{irsf}. From the source density of the field we calculate the probability of a chance association between the \chan\ transient and the infrared counterpart to be approximately 1 percent.

\begin{table}
\centering
\caption{Magnitudes in the {\it J}, {\it H} and {\it K$_s$} bands for the transient XRT~120830.}
\begin{tabular}{cccc}
\hline
Telescope & {\it J}-band & {\it H}-band & {\it K$_s$}-band \\
\hline
2MASS & 16.8$\pm$0.2 & 15.9$\pm$0.2 & 15.5$\pm$0.2 \\
IRSF/SIRIUS & 16.70$\pm$0.06 & 15.92$\pm$0.04 & 15.37$\pm$0.06 \\
\hline
\end{tabular}
\label{irsf_mag}
\end{table}

We have fitted a powerlaw model (pegpwrlw in Xspec) to the X-ray spectrum of XRT~120830. As the proper motion of the possible counterpart suggests this is a nearby source we apply no correction for absorption. Due to the low count rate we have taken source photons from the start of the burst to the end of the observation to get as many counts as possible. We find a photon index of $2.5\pm0.2$ and a peak flux of $(8.7\pm0.7)\times10^{-11}$ \flx. We do not find a sufficient count rate to follow the spectral evolution of the source with a hardness ratio.

\section{Discussion}
\subsection{XRT~110103}

Without a counterpart and a subsequent redshift measurement it is not clear whether XRT~110103 is at the distance of the galaxy cluster ACO 3581 or originates from a foreground or even more distant object. If it is associated with the cluster ACO 3581 that would place it at a distance of 94.9 Mpc \citep{johnstone05}, impying a peak luminosity of $\sim2\times10^{44}$ \lum. The Galactic latitude of XRT~110103 is $b\sim32.7$ suggesting a possible Galactic source is unlikely to be further than $\sim1$ kpc. At this distance, the peak luminosity would be $\sim2\times10^{34}$ \lum which suggests an X-ray burst is unlikely \citep{galloway08}.

XRT~110103 exhibits similar properties to the transient XRT~000519 reported by \citet{jonker13}. Both evolve over a similar time-scale with the main burst taking place over a few hundred seconds and a slow powerlaw decay over the remainder of each observation. The lightcurve of XRT~000519 showed a twin peak in the main flare which is not evident in XRT~110103. Both have similar powerlaw spectra and were detected in observations of galaxy clusters. However, we are unable to associate XRT~110103 with any known counterpart within the cluster ACO 3581.  XRT~000519 displayed precursor events $\sim4$ and $8$ ks before the main flare. In the 6 ks of observations prior to the main flare of XRT~110103 we find no precursor events. XRT~110103 is a factor of a few fainter in flux so we cannot rule out the possibility that similar precursor events are present but below our detection threshold. Jonker et al.~(2013) offer multiple explanations for XRT~000519, favouring the tidal disruption of a white dwarf by an intermediate mass black hole. They do not rule out alternative scenarios such as the accretion of an asteroid by a foreground neutron star or an off-axis $\gamma$-ray burst (GRB).

\citet{jonker13} estimate a peak luminosity for XRT~000519 of $\sim6\times10^{42}$ \lum assuming a transient distance of 16.2 Mpc, equivalent to that of M86. If these events are due to the tidal disruption of a white dwarf by an intermediate mass black hole the different peak luminosities could be due to different black hole masses or due to different beaming factors \citep[e.g.][]{macleod14}. The transient recently discovered in the Chandra Deep Field-South survey \citep{atel6541} could also be related to XRT~110103 and XRT~000519. The count rate quoted by \citet{atel6541} and redshift for a potential counterpart galaxy of 0.31 \citep{atel6603} suggest a peak luminosity of $\sim4\times10^{44}$ \lum.

These events could also be interpreted as X-ray flashes, which are probably related to GRBs \citep[e.g.][]{sakamoto05}. The time-scale of the main flares and the late-time powerlaw decays seen in XRT~110103 and XRT~000519 are consistent with the early X-ray emission typically seen in GRB lightcurves \citep[e.g.][]{obrien06}. Figure 7 in \citet{sakamoto05} shows the peak energy constraints on X-ray flashes, X-ray-rich GRBs and Hard GRBs. Of these scenarios, the upper limits on the peak energy of XRT~110103 and XRT~000519 from the power law fits could only be consistent with an X-ray flash. However, for XRT~000519 the Burst and Transient Source Experiment (BATSE) detected no gamma-ray burst at the time of the X-ray transient \citep{jonker13}.

If XRT~110103 and XRT~000519 are the same type of object we provide an estimate of the rates for these events. We searched a total of 8.4 years of ACIS-I and ACIS-S public data with no grating used and found two events of this type. Assuming an approximate 16 $\times $ 16 arc-minute field of view we derive a rate of $1.4\times10^{5}$ per year over the whole sky with a peak X-ray flux greater than $10^{-10}$ \flx. There are significant biases in the sky sampling from \chan\ observations. If these transients are extragalactic then X-ray absorption in any Galactic plane observations would likely result in any such transients being obscured. Further if these events are associated with galaxy clusters then the \chan\ observations devoted to clusters (approximately 18.5 per cent of ACIS observations are listed in this science category) would have a greater chance of observing such an event than a random location on the sky. Without firm knowledge of the nature or distance of these transients it is difficult to account for these possible biases. This combined with the small sample size suggests this rate should be treated as a very rough approximation. As we have two similar transients with such a large flux there should be many more fainter transients of a similar nature in the \chan\ archive. There should also be events in the archives of other soft X-ray telescopes such as {\it ROSAT}. Assuming a lightcurve and model similar to the best fit we find for XRT~110103 and a log-N log-S slope of -1.5 we calculate that there should be as many as 100 such events in the ROSAT All Sky Survey with greater than 15 photons detected. \citet{greiner00} may have found some of these events.

\subsection{XRT~120830}

XRT~120830 has a Galactic latitude of $b\sim-67.2$ which suggests a Galactic source would be within $\sim1$ kpc. At this distance we find a peak luminosity of $\sim10^{34}$ \lum which again makes an X-ray burst unlikely.

The proper motion of $74\pm19$ milliarcsec per year for the likely counterpart to XRT~120830 suggests a flare from a nearby dwarf star as the most likely explanation for the X-ray transient. If we assume a typical tangential velocity for a local dwarf star of $\sim25$ km s\textsuperscript{-1} (e.g. see figure 7 of \citealt{faherty09}) we find a distance of $\sim80$ pc. The absolute magnitude of $M_J\sim12.2$ at such a distance would imply a dwarf star around the L1 spectral type \citep[see equation 6 and figure 7 in][]{cruz03}. The infrared colours are also consistent with a late M or early L dwarf \citep[e.g. figure 6 in][]{cruz03}, suggesting that this is plausible distance. At this distance we estimate a peak X-ray luminosity of $\sim7\times10^{31}$ \lum. Larger peak X-ray luminosities have been reported in superflares for earlier type M dwarfs \citep[e.g.][]{osten10}. This may be the largest X-ray flare detected in such a late type dwarf, although the distance and spectral type are uncertain. Spectroscopic followup is required to confirm both the spectral type and provide a more accurate distance estimate.

The marginal flaring events 7 and 14 ks after the main flare could be related to the rotation period of the dwarf star in this case, although we cannot draw any firm conclusions regarding periodicity from 3 events. A $\sim2$ hour rotation period would be rapid for an M dwarf \citep[e.g.][]{delfosse98} but some late-type dwarfs have comparable periods such as BRI 0021-0214 at $\sim2.5$ hours \citep{basri95} and 2MASSW J1707183+643933 at 3.6 hours which has also displayed a strong flare with an amplitude in the UV of at least 6 mag \citep{rockenfeller06}.

\section{Conclusions}

In summary, we have discovered two previously unknown transients in archival \chan\ data. One of these events, XRT~110103, could be extra-galactic in nature and deeper optical follow-up of XRT~110103 should help to determine the nature of this transient. Combined with XRT~000519 and the transient reported by \citet{atel6541} there is a population of \chan\ transients which could be of similar nature. A wider sample of such objects, from the \chan\ archive or other soft X-ray telescopes, should help to determine whether these events are related to a known class of transient.

XRT~120830 appears to be consistent with an X-ray flare from a late M or early L dwarf star. We are unable to confirm the precise spectral type of the dwarf from the infrared colours and magnitudes. The possible minor flares $\sim7$ and $14$ ks after the main flare could be related to the rotation period of the star.

Whilst telescopes with a large field of view such as {\it Swift-BAT} are ideal for detecting large numbers of transients those discovered serendipitously during routine observations can provide unique information. Soft X-ray lightcurves and spectra are available for the entire transient event in this case rather than being limited to the afterglow. In the case of XRT~000519 the availability of a lightcurve prior to the main flare revealed possible precursor events. However, discovery years after the event means followup observations are likely to miss the decay phase of most transients. Some form of quick alert system could rectify this, even if only alerting the principal investigator of the observation if data exclusivity periods must be adhered to.

\section*{Acknowledgements}

A.G. acknowledges the assistance of Patrick Woudt in obtaining the IRSF/SIRIUS observations used in this paper.

This project was funded in part by European Research Council Advanced Grant 267697 '4 pi sky: Extreme Astrophysics with Revolutionary Radio Telescopes.'

This publication makes use of data products from the Two Micron All Sky Survey, which is a joint project of the University of Massachusetts and the Infrared Processing and Analysis Center/California Institute of Technology, funded by the National Aeronautics and Space Administration and the National Science Foundation.

This publication makes use of data products from the Wide-field Infrared Survey Explorer, which is a joint project of the University of California, Los Angeles, and the Jet Propulsion Laboratory/California Institute of Technology, funded by the National Aeronautics and Space Administration.

This research has made use of the USNOFS Image and Catalogue Archive operated by the United States Naval Observatory, Flagstaff Station (http://www.nofs.navy.mil/data/fchpix/).


\begin{thebibliography}{99}
\bibitem[\protect\citeauthoryear{Basri \& Marcy}{1995}]{basri95} Basri G., Marcy G.~W., 1995, AJ, 109, 762 
\bibitem[\protect\citeauthoryear{Bessell \& Brett}{1988}]{bessell88} Bessell M.~S., Brett J.~M., 1988, PASP, 100, 1134 
\bibitem[\protect\citeauthoryear{Carpenter}{2001}]{carpenter01} Carpenter J.~M., 2001, AJ, 121, 2851 
\bibitem[\protect\citeauthoryear{Cash}{1979}]{cash79} Cash W., 1979, ApJ, 228, 939 
\bibitem[\protect\citeauthoryear{Cruz et al.}{2003}]{cruz03} Cruz K.~L., Reid I.~N., Liebert J., Kirkpatrick J.~D., Lowrance P.~J., 2003, AJ, 126, 2421
\bibitem[\protect\citeauthoryear{Cutri et al.}{2012}]{cutri12} Cutri R.~M., et al., 2012, wise.rept, 1
\bibitem[\protect\citeauthoryear{Delfosse et al.}{1998}]{delfosse98} Delfosse X., Forveille T., Perrier C., Mayor M., 1998, A\&A, 331, 581 
\bibitem[\protect\citeauthoryear{Dickey \& Lockman}{1990}]{dickey90} Dickey J.~M., Lockman F.~J., 1990, ARA\&A, 28, 215 
\bibitem[\protect\citeauthoryear{Evans et al.}{2010}]{evans10} Evans I.~N., et al., 2010, ApJS, 189, 37 
\bibitem[\protect\citeauthoryear{Faherty et al.}{2009}]{faherty09} Faherty J.~K., Burgasser A.~J., Cruz K.~L., Shara M.~M., Walter F.~M., Gelino C.~R., 2009, AJ, 137, 1 
\bibitem[\protect\citeauthoryear{Galloway et al.}{2008}]{galloway08} Galloway D.~K., Muno M.~P., Hartman J.~M., Psaltis D., Chakrabarty D., 2008, ApJS, 179, 360 
\bibitem[\protect\citeauthoryear{Greiner et al.}{2000}]{greiner00} Greiner J., Hartmann D.~H., Voges W., Boller T., Schwarz R., Zharikov S.~V., 2000, A\&A, 353, 998 
\bibitem[\protect\citeauthoryear{Johnstone et al.}{2005}]{johnstone05} Johnstone R.~M., Fabian A.~C., Morris R.~G., Taylor G.~B., 2005, MNRAS, 356, 237 
\bibitem[\protect\citeauthoryear{Jonker et al.}{2013}]{jonker13} Jonker P.~G., et al., 2013, ApJ, 779, 14 
\bibitem[\protect\citeauthoryear{Kato et al.}{2007}]{kato07} Kato D., et al., 2007, PASJ, 59, 615 
\bibitem[\protect\citeauthoryear{Kim et al.}{2007}]{kim07} Kim M., et al., 2007, ApJS, 169, 401 
\bibitem[\protect\citeauthoryear{Komossa et al.}{2004}]{komossa04} Komossa S., Halpern J., Schartel N., Hasinger G., Santos-Lleo M., Predehl P., 2004, ApJ, 603, L17
\bibitem[\protect\citeauthoryear{Lewin \& van der Klis}{2006}]{lewin06} Lewin W.~H.~G., van der Klis M., 2006, csxs.book,
\bibitem[\protect\citeauthoryear{Lodato \& Rossi}{2011}]{lodato11} Lodato G., Rossi E.~M., 2011, MNRAS, 410, 359
\bibitem[\protect\citeauthoryear{Luo, Brandt, \& Bauer}{2014}]{atel6541} Luo B., Brandt N., Bauer F., 2014, ATel, 6541, 1
\bibitem[\protect\citeauthoryear{MacLeod et al.}{2014}]{macleod14} MacLeod M., Goldstein J., Ramirez-Ruiz E., Guillochon J., Samsing J., 2014, ApJ, 794, 9 
\bibitem[\protect\citeauthoryear{Monet et al.}{2003}]{monet03} Monet D.~G., et al., 2003, AJ, 125, 984
\bibitem[Nagayama et al.(2003)]{nagayama03} Nagayama, T., Nagashima, C., Nakajima, Y., et al.\ 2003, \procspie, 4841, 459 
\bibitem[\protect\citeauthoryear{O'Brien et al.}{2006}]{obrien06} O'Brien P.~T., et al., 2006, ApJ, 647, 1213
\bibitem[\protect\citeauthoryear{Osten et al.}{2010}]{osten10} Osten R.~A., et al., 2010, ApJ, 721, 785 
\bibitem[\protect\citeauthoryear{Osten et al.}{2005}]{osten05} Osten R.~A., Hawley S.~L., Allred J.~C., Johns-Krull C.~M., Roark C., 2005, ApJ, 621, 398 
\bibitem[\protect\citeauthoryear{Rockenfeller et al.}{2006}]{rockenfeller06} Rockenfeller B., Bailer-Jones C.~A.~L., Mundt R., Ibrahimov M.~A., 2006, MNRAS, 367, 407
\bibitem[\protect\citeauthoryear{Rosswog, Ramirez-Ruiz, \& Hix}{2009}]{rosswog09} Rosswog S., Ramirez-Ruiz E., Hix W.~R., 2009, ApJ, 695, 404 
\bibitem[\protect\citeauthoryear{Rots \& Budav{\'a}ri}{2011}]{rots11} Rots A.~H., Budav{\'a}ri T., 2011, ApJS, 192, 8 
\bibitem[\protect\citeauthoryear{Sakamoto et al.}{2008}]{sakamoto08} Sakamoto T., et al., 2008, ApJ, 679, 570 
\bibitem[\protect\citeauthoryear{Sakamoto et al.}{2005}]{sakamoto05} Sakamoto T., et al., 2005, ApJ, 629, 311
\bibitem[\protect\citeauthoryear{Sidoli}{2013}]{sidoli13} Sidoli L., 2013, arXiv, arXiv:1301.7574 
\bibitem[\protect\citeauthoryear{Skrutskie et al.}{2006}]{skrutskie06} Skrutskie M.~F., et al., 2006, AJ, 131, 1163 
\bibitem[\protect\citeauthoryear{Strohmayer \& Bildsten}{2006}]{strohmayer06} Strohmayer T., Bildsten L., 2006, csxs.book, 113
\bibitem[\protect\citeauthoryear{Treister, Bauer, \& Schawinski}{2014}]{atel6603} Treister E., Bauer F., Schawinski K., 2014, ATel, 6603, 1 
\end{thebibliography}
\end{document}